\documentclass[%
reprint,
twocolumn,
superscriptaddress,showkeys,
 amsmath,amssymb,
 aps,
prl,
]{revtex4-1}
\usepackage{epstopdf}
\usepackage{amssymb}
\usepackage{color}
\usepackage{graphicx}
\usepackage{dcolumn}
\usepackage{bm}
\usepackage{hyperref}

\usepackage[latin1]{inputenc}
\usepackage{amsmath}
\usepackage{amsfonts}
\usepackage{amssymb}
\usepackage{graphicx}
\usepackage{verbatim}
\begin{document}

\preprint{}
\title{Excited to excited state scattering using weak measurements.}

\author{Satya Sainadh U}
\affiliation{Center for Quantum Dynamics, Griffith University, Brisbane, QLD 4111, Australia}
\author{Andal Narayanan}%
 \email{andal@rri.res.in}
\affiliation{%
 Light and Matter Physics Group,
 Raman Research Institute, Bangalore 560080, India
}%



\date{\today}

\begin{abstract}
Weak measurements are a subset of measurement processes in quantum mechanics wherein the system 
which is being measured interacts very weakly with the measuring apparatus.~Measurement values of observables
undergoing a weak interaction and their amplification, are concepts that have sharpened our understanding
of interaction processes in quantum mechanics.~Recent experiments show that naturally occurring processes such as 
resonance fluorescence from excited states of an atom can exhibit weak value amplification effect.
In this paper, we theoretically analyze the process of elastic resonance fluorescence from a V-type three level atomic
system, using the well known Weiskopff-Wigner (W-W) theory of spontaneous emission.~Within this theory, we show that, 
a weak interaction regime can be identified and for suitable choices of initial and final excited states, 
the mean scattering time between these states show an amplification effect during interaction with the vacuum bath modes
of the electromagnetic field. We thus show for the first time that a system-bath interaction can show weak value amplification.
Using our theory we reproduce the published experimental results carried out in such a system. More importantly, our theory can
calculate scattering timescales in elastic resonance scattering between multiple excited states of a single atom or 
between common excited state configurations of interacting multi-atom systems.
\end{abstract}

\keywords{Weak interaction and weak value amplification, Resonance fluorescence}
\maketitle

\section{Introduction.}
\begin{figure}
\includegraphics[scale=0.45]{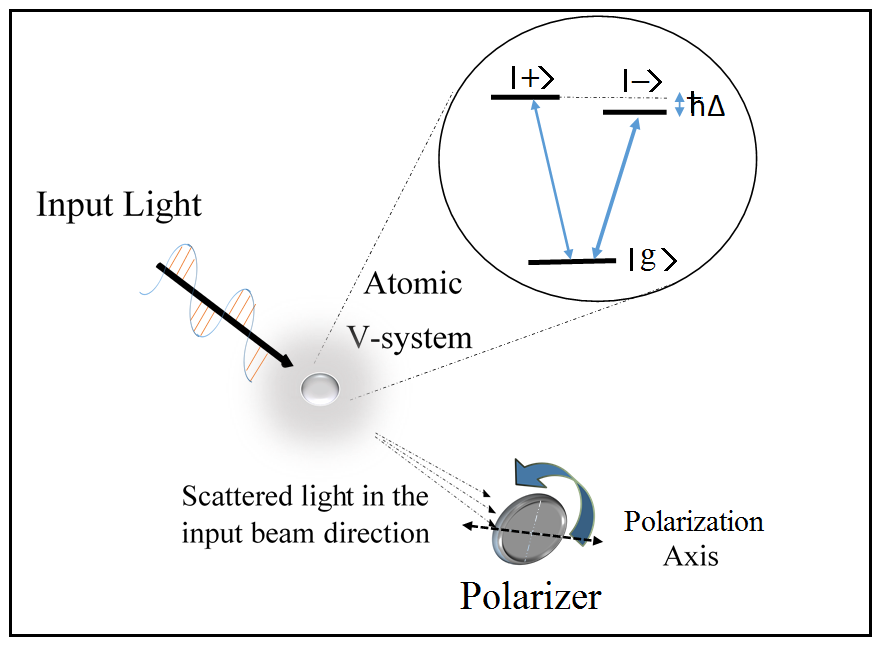}
\caption{\label{fig1}(Color online) An atomic V level configuration with ground state $\vert g \rangle$ and two excited states $\vert + \rangle$ and 
$\vert - \rangle$.~The separation between the excited states $\Delta$ is less than $\Gamma$, the rate
of excited state decay.~The elastically scattered light in the forward direction is detected for a particular polarization orientation.}
\end{figure}

Weak measurements refer to measurement processes in quantum mechanics, where the measuring device has very little effect on the system which
is being measured.~Weak interaction is a more general concept, 
referring to an interaction process where the strength of coupling between the 
observables taking part in the interaction is very weak.~Weak values of an observable 
is a relatively new concept that is definable in the case of weak measurements on the observable that has undergone a
weak interaction.
This concept was first introduced by Aharonov, Albert and Vaidman
 \cite{aharanov-prl-60-1351-1988}.~They showed that the weak value associated with an observable undergoing weak interaction, 
can exhibit an amplification effect called weak value amplification. This effect occurs 
for pre and post selected quantum states of the observable that are nearly orthogonal to each other.~It was pointed out subsequently  \cite{sudarshan-prd-40-2112-1989} that, indeed under suitable conditions,
a projective post-selection of the outcome of a weak interaction  
can substantially alter the value of the measured observable, from its eigen value measurements, which is obtained
without post-selection.~It is argued that, the advantages of weak value amplification effect can be leveraged when the measurement process
is dominated by technical noise and not by the shot noise of the detector
 \cite{simon-prl-105-010405-2010, steinberg-prl-107-133603-2011}.\\
\indent Experiments demonstrating weak measurements and weak interactions have been performed.~Using weak interaction in an Young's double slit experimental set-up,
the average trajectories of photons exiting either one or other of the two slits, have been mapped
  \cite{steinberg-science-332-1170-2011} without loss in the visibility of interference fringes.~Since weak values can even be complex, it has enabled direct measurement of
complex probability amplitudes as is shown in  \cite{lundeen-nat-474-188-2011,boyd-natpho-7-316-2013}.~Experiments
are also done to bring out the weak value amplification effect.~These experiments  \cite{kwait-science-319-787-2008,prl-102-173601-2009}
measured effects, using weak value amplification, which would have otherwise been too small to be measured.~However, 
many of these experiments have been deliberately
designed to bring out the effects of weak value amplification of the observable being measured.\\
\indent By contrast in a recent experiment  \cite{dayan-prl-111-023604-2013}, Shomroni et.~al measured an imaginary weak value
which is the arrival time of photons during a natural decay processes of an excited state atom.~They chose a
V-type three level atomic configuration initially prepared in a superposition of excited levels to be their system.This interacts
with the vacuum electromagnetic bath environment (See Fig. \ref{fig1}).
They showed that for appropriately selected initial and final states of the system, the observable being measured showed weak value amplification,
which manifested as a delayed arrival of photon as compared to the mean arrival time of photons from an excited state decay to the ground state.
\\
\indent In this paper, we apply the well known Weiskopff-Wigner (W-W) theory of spontaneous emission for an atom in the excited states of a V-type
configuration of levels. Using this theory we derive mean scattering time of the atom between
two well defined excited state configurations.~By our calculation, we explicitly show that the extended
excited state lifetime seen in the experiment of \cite{dayan-prl-111-023604-2013}, is actually the mean scattering time from a particular, 
initial superposition state of the two excited states of the V-system, to a nearly orthogonal post-selected final excited state 
superposition.~It is important to 
note that the atom does not decay to the ground state during this scattering event and thus we are in the elastic
scattering regime of resonance fluorescence.~Outside of this regime and without post-selection, the scattering will be inelastic, dominated
by excited state decay to the ground state.~We thus show for the first time 
that elastic resonance fluorescence from multiple excited states modeled by W-W theory can be used to describe a weak interaction
in these systems.~More importantly, we show using W-W theory that
weak value amplification occurs naturally for configurational changes between excited states which are nearly orthogonal to each other during
interaction with vacuum bath modes of the EM field.
Such scattering events are rare, before which, in a usual measurement process of atomic decay, the atom would have decayed to the ground state.
The weak value amplification process serves as a time sieve separating the rare excited state configurational changes
from the overwhelming frequent decay of the excited state to the ground state.~It thus allows
for real-time experimental observation of excited state configurational changes from a collection of events which overwhelmingly
contain excited state decay to the ground state.
\section{Resonance fluorescence from a V-system.}
A V-type atomic level configuration is a prototypical quantum atom-optical system.~Initial theoretical 
studies on resonance fluorescence from this system 
\citep{dalibard-epl-1-441,pra-46-373} addressed the origin of the light and dark periods seen in 
their resonance fluorescence \citep{prl-56-2797}.~These features were explained on the basis of formation of symmetric 
$\vert S \rangle = \frac{\vert + \rangle + \vert - \rangle}{\sqrt{2}}$ and anti-symmetric 
$\vert A \rangle = \frac{\vert + \rangle - \vert - \rangle}{\sqrt{2}}$ combination of excited states.~At any given time, for finite detuning,
the state of the atom will be in a mixture of $\vert S \rangle$ and $\vert A \rangle$ states.~The symmetric state $\vert S \rangle$, couples 
to the vacuum modes of electromagnetic bath field and rapidly de-excites to the ground state.~Before such a decay can happen, there is a finite but small
probability for the $\vert S \rangle$ state to get converted to the $\vert A \rangle$ state which does not couple to the field.~This process
combined with the depleting population of the $\vert S \rangle$, gives rise to dark periods in the fluorescence spectrum.~This effect happens 
independent of the relative orientation of induced dipoles between the ground state $\vert g \rangle$ 
and the excited states $\vert + \rangle$ and $\vert - \rangle$.~For parallel orientation of the induced dipoles, the $\vert A \rangle$ state
becomes invariant in time and a total vanishing of fluorescence due to quantum interference was predicted  \cite{pra-46-373}.~\\
\indent It is clear from the above analysis that, for non-parallel orientations of induced dipoles, the rate at which the $\vert S \rangle$
state changes to $\vert A \rangle$ state without de-excitation is the regime of elastic resonance where the incident light on such a 
system undergoes merely a phase change in the forward scattering direction  \cite{mollow-phys-rev-188-1969-1969}.
\section{Weak value amplification in resonance fluorescence}
We now briefly outline the traditional weak value amplification theory.
The weak value of an observable is defined as  \cite{aharanov-prl-60-1351-1988},
\small\begin{equation}
\langle O_w \rangle = \frac{\langle f| \hat{O} | i \rangle}{\langle f|i \rangle}  .
\end{equation}\normalsize
It is clear from the above definition that for a non zero 
$\langle f| \hat{O} | i \rangle$, the weak value $O_w$ gets amplified for nearly orthogonal $\vert i \rangle$ and $\vert f \rangle$ states and that the 
weak value can lie outside the eigenvalue value spectrum of the observable $\hat{O}$.\\
\indent For elastic scattering in resonance fluorescence from the V-system the observable $\hat{O}$, which takes an initial symmetric atomic state 
$\vert i \rangle = \vert S \rangle$ of the atom to a nearly orthogonal final state,
$
\vert f \rangle = \frac{1}{\sqrt{2}}\left(e^{-i \epsilon }|+\rangle -e^{+i \epsilon} |-\rangle\right)$ where $\epsilon\ll1$, 
is the $\sigma_z = \vert + \rangle \langle + \vert - \vert -\rangle \langle - \vert$ observable.~Here the value of $\epsilon$ lies between 0 and 1.
As is shown in the experiment of  
 \cite{dayan-prl-111-023604-2013}, evaluating the weak value of this $\sigma_z$ observable between nearly orthogonal $\vert i \rangle$
and $\vert f \rangle$ states produces a weak value amplification effect.~This manifests as an increase in the arrival
time of photons on a detector which detects only the $\vert f \rangle$ (post-selection) state.~Based on our physical picture of elastic scattering
between the $\vert S \rangle$ and $\vert A \rangle$ states, we explain this increased mean life-time as the mean scattering timescale
between the $\vert i \rangle$ and $\vert f \rangle$ states using the well known Weiskopff-Wigner theory of spontaneous emission.
\section{Weiskopff-Wigner theory for resonance fluorescence from a V-system.}
Before proceeding to calculate the mean scattering time between chosen states, we estimate the time-scales inherent in the problem.
The time-scale of atomic evolution between $\vert S \rangle$ and $\vert A \rangle$ states without decay, depends on the energy difference between
$\vert + \rangle$ and $\vert - \rangle$ states and hence proportional to $1/\Delta$ (See Fig. \ref{fig1}).~The larger the value of $\Delta$ 
the faster is this evolution.~On the other hand, 
the mean time-scale of decay to the ground state from either of the excited state is given by $1/\Gamma$.~
In order that the scattering is in the elastic regime, we need to satisfy {\color{black} $\frac{1}{\Gamma} \ll \frac{1}{\Delta}$
i.e., $\Delta \ll \Gamma$}.~We wish to
point out that this is precisely the regime where the experiment reported in  \cite{dayan-prl-111-023604-2013} was conducted.~This regime naturally
emerges for our calculation if we demand that the scattering between the $\vert i \rangle$
and the $\vert f \rangle$ states be an elastic scattering.\\
\indent{\color{black}  We now proceed to calculate the probability that an atom, initially in the state $\vert i \rangle$ gets
scattered to a nearly orthogonal final state $\vert f \rangle$ during its interaction with the vacuum modes. Our calculation is based on
the well known Weiskopff-Wigner derivation originally proposed for a two-level excited state atom interacting with vacuum bath modes
of the EM field.}\\
\indent The state of the atom at any time $t$ is given by
\small\begin{eqnarray}
\vert \psi(t) \rangle &=& \alpha(t)  e^{-iwt}\vert \psi_{in},0_{k,s}\rangle+ 
                      \sum_{{\bf k},s} \beta^1_{{\bf k},s}(t)e^{-iw_{k}t}\vert g,1_{k,s}\rangle  \nonumber \\
            & +& \sum_{{\bf k'},s'}\beta^2_{{\bf k'},s'}(t)e^{-iw_{k'}t}\vert g,1_{k',s'}\rangle .\label{eq:psit} 
\end{eqnarray}\normalsize
Here $\alpha(t)$ refers to the probability amplitude that the atom is in a particular combination of the excited states $|+\rangle$ and $|-\rangle$
represented by $\vert \psi_{in} \rangle$ 
with no photons.~The energy of this state is $\hbar w$.~The terms $\beta_{k,s}^{1,2}$ refer to the probability amplitudes that the atom decays 
to the ground state  $\vert g \rangle$ from either of the excited states, emitting a photon.~The appropriate mode of the emitted photon is labeled by
the propagation vector of the emitted photon ${\bf k}$ or ${\bf k'}$, its polarization state $s$ or $s'$ and its energy $w_k$ or $w_k'$ 
distinguishing emission from two different excited states to the ground state.
We take the atom to be initially in the state $\vert i \rangle = \vert S \rangle$.~
The Hamiltonian of this atom interacting with the electromagnetic vacuum bath
modes is given by
\small
\begin{eqnarray}
\hat{H}/\hbar&=& w_+ \sigma_{e_+ e_+}+w_-\sigma_{e_- e_-}+\sum_{k,s}w_k
\hat{n}_{k,s}\nonumber\\&-&\sum_{k,s}g^1_{k,s}\sigma_{e_+g}\hat{a}_{k,s}-\sum_{k',s'}g^2_{k',s'}
\sigma_{e_-g}\hat{a}_{k',s'}+h.c.~.\label{eq:hamil}
\end{eqnarray}
\normalsize
Here $\sigma_{e{\pm}} \equiv \vert e_{\pm} \rangle \langle g \vert $  is the operator representing transition from either of the excited states 
$e_{\pm}$ to the ground state.~The coupling constant of excited states to the vacuum modes
is, $g_{k,s}=-i\sqrt{\frac{w_k}{2\hbar\varepsilon_0V}}(\mathbf{d.\epsilon_{k,s}}) $ where 
$V$ is the quantization volume and $\mathbf{d}$ is the induced dipole moment 
vector for the transition and  $\mathbf{\epsilon_{k,s}}$ is the unit vector along a polarization $s$.
Using {\color{black} Schroedinger Wave Equation (SWE), we get the evolution equations for $\alpha$ and both the $\beta$s.}
\small\begin{eqnarray}
H\vert \psi(t) \rangle &=&i\hbar \frac{\partial \psi}{\partial t}, \label{eq:hamil} \label{eq:ampli}\\
&=& i\hbar \left(  \dot{\alpha}(t)-i w \alpha(t)\right)e^{-iw t} \vert \psi_{in},0_{k,s} \rangle +  \nonumber\\
& & i \hbar \left( \dot{\beta}^1_{k,s} (t)-iw_k\beta^1_{k,s}(t)\right)e^{-iw_kt} \vert g,1_{k,s} \rangle  +  \nonumber \\
& & i \hbar \left( \dot{\beta}^2_{k',s'} (t)-iw_{k'}\beta^2_{k',s'}(t)\right)e^{-iw_{k't}} \vert g,1_{k',s'} \rangle .
\end{eqnarray}\normalsize
The LHS of equation (\ref{eq:ampli}) can be calculated using the definitions of $H$ and $\vert \psi(t) \rangle$.\\
\indent The problem we are addressing pertains to scattering between two different configurations of the excited states denoted by $\vert i \rangle$
and $\vert f \rangle$.~Therefore using equations(\ref{eq:psit})-(\ref{eq:hamil}) and using
the  Schroedinger wave equation (SWE) $H\vert \psi(t) \rangle =i\hbar \frac{\partial \vert \psi(t) \rangle}{\partial t}$ we take the inner product 
of SWE with $\langle f \vert$ , $\langle g,1_{k,s}\vert $ and $\langle g,1_{k',s'}\vert $ separately on both sides.~This is 
equivalent to post-selection on a particular outcome.~{\color{black} Since our desired final state $\vert f \rangle$ does not involve the ground state,
this post-selection represents a scattering event during which the atom in the excited state $\vert i \rangle$ does not decay to the ground state.}
{\color{black} After post-selection the dynamical equations representing the probability amplitudes are given as}
\small
\begin{eqnarray}
\dot{\alpha}(t)&=&-i\left( \frac{w_++w_-}{2}-w\right)\alpha(t) + \frac{\Delta}{2} \text{cot}(\epsilon) \alpha(t) +  \\
& &\frac{i}{\sqrt{2}}\sum_{k,s}\left(\beta^1_{k,s}(t)+\beta^2_{k,s}(t)\right)\left( g^1_{k,s}-g^2_{k,s}\right)e^{-i(w_k-w)t}, \nonumber \\
\dot{\beta}^1_{k,s}(t)&=& \frac{i\alpha(t)}{\sqrt{2}}\left( g^{1*}_{k,s}-g^{2*}_{k,s}\right)e^{i(w_k-w)t}-iw_k \beta^2_{k,s}(t), \nonumber \\
\dot{\beta}^2_{k',s'}(t)&=& \frac{i\alpha(t)}{\sqrt{2}}\left( g^{1*}_{k',s'}-g^{2*}_{k',s'}\right)e^{i(w_{k'}-w)t}-iw_{k'} \beta^1_{k',s'}(t) .
\end{eqnarray}\normalsize
By defining $\beta^1_{k,s}(t)+\beta^2_{k,s}(t)=\beta_{k,s}(t)$  and $g^1_{k,s}-g^2_{k,s}=g_{k,s}$ and by using the fact that the separation
between the excited states $\Delta$ is less than that of the width of their energy levels i.e.~$\Delta \ll \Gamma$, we assume $w_+ \approx w_-$.
{\color{black} This translates to the condition that in the weak interaction regime the energies of both the excited states of the V system are almost
equal}, thereby making $w =(w_1+w_2)/2$.~Using this approximation, we obtain the following equations:
\small
\begin{eqnarray}
\dot{\alpha}(t)&=&\frac{i}{\sqrt{2}}\sum_{k,s}\beta_{k,s}(t) g_{k,s}e^{-i(w_k-w)t} + \frac{\Delta}{2} \text{cot}(\epsilon) \alpha(t), \label{eq:aoft}\\
\dot{\beta}_{k,s}(t)&=& \sqrt{2}i\alpha(t) g^{*}_{k,s}e^{i(w_k-w_3)t}-iw_k \beta_{k,s}(t). \label{eq:boft}
\end{eqnarray}\normalsize
The formal solution for $\beta_{k,s}(t)$ is \small
\begin{eqnarray}
\beta_{k,s}(t)=e^{-w_kt}\beta_{k,s}(0)+i \sqrt{2} g^{*}_{k,s}\int_0^t\alpha(t')e^{i(w_k-w_3)t'}dt'.
\end{eqnarray}\normalsize
Since at time $t=0$ we consider all the atoms to occupy the excited state configuration $\vert i \rangle$ {\color{black} the initial value of}
$\beta_{k,s}(0)=0$. We then obtain,\small
\begin{equation}
\dot{\alpha}(t)=-\sum_{k,s}|g_{k,s}|^2\int_0^t\alpha(t')e^{i(w_k-w)(t-t')}dt' +\frac{\Delta}{2} \text{cot}(\epsilon) \alpha(t).
\end{equation}\normalsize
{\color{black} Since the emitted photon can be in any direction $k$ with any polarization $s$, we have to sum over these two variables.}
In the continuum limit (i.e., when the quantization volume $V \rightarrow \infty$) 
\small \begin{equation*}
\sum_{k,s} \simeq\sum_{s=1}^2\int\mathcal{D}(k)d^3k,
\end{equation*}\normalsize where $\mathcal{D}(k)$ 
is the density of states in $k$-space.~Using spherical coordinates in $k$ space denoted by $(k, \theta, \phi)$, 
we have \small\begin{equation*}\int\mathcal{D}(k)d^3k=\frac{V}{(2 \pi) ^3}\int_0^\infty k^2dk\int_0^\pi \text{sin} \theta d\theta \int_0^{2\pi}d\phi. \end{equation*}\normalsize
Now by using the definition of $g_{k,s}$ we evaluate\small
\begin{equation}
\sum_{k,s}|g_{k,s}|^2=\int_0^\infty \frac{w_k}{2(2 \pi) ^3\varepsilon_0\hbar}k^2dk\sum_{s=1}^2\left[\int_0^\pi \text{sin}
 \theta d\theta \int_0^{2\pi}(\mathbf{d.\epsilon_{k,s}})^2d\phi\right]. 
\end{equation}\normalsize
Since the triplet $(\mathbf{\epsilon_{k,1},\epsilon_{k,2},\kappa})$  with $\mathbf{\kappa}=\mathbf{k}/k$ forms an orthogonal coordinate system,
\small\begin{eqnarray}
&&\mathbf{d}=(\mathbf{d.\epsilon_{k,1}})\epsilon_{k,1}+(\mathbf{d.\epsilon_{k,2}})\epsilon_{k,2}+(\mathbf{d.\kappa })\mathbf{\kappa}, \\
\therefore &&\sum_{s=1}^2\mathbf{(d.\epsilon_{k,s}})^2 = \mathbf{|d|^2}-\mathbf{|d.\kappa|^2 }.
\end{eqnarray}\normalsize  
For the transitions where $\Delta m=\pm 1$ the induced dipole moment vectors are complex and {\color{black} orthogonal to each other}.~By defining  $\mathbf{d=d_1-d_2}$ and considering the case of these two 
induced dipole moment vectors being perpendicular to each other {\color{black} we define without loss of generality}
\small \begin{equation*}
\mathbf{d_1}=|\eta|\frac{\mathbf{x}+i\mathbf{y}}{\sqrt{2}}, \qquad 
\mathbf{d_2}=|\eta|\frac{\mathbf{x}-i\mathbf{y}}{\sqrt{2}}.
\end{equation*}\normalsize
Therefore  $\mathbf{d=d_1-d_2}=i\sqrt{2}|\eta|\mathbf{y}$.~Incorporating {\color{black} this in the evolution equations}, we get\small 
\begin{eqnarray}
\dot{\alpha}(t)&=&-\sum_{k,s}|g_{k,s}|^2\int_0^t \alpha(t')e^{i(w_k-w)(t-t')}dt', \label{eq:finala} \\ 
&=& \frac{|\eta|^2}{6\pi ^2\hbar \varepsilon_0c^3}\int_0^\infty w_k^3dw_k\int_0^t \alpha(t')e^{i(w_k-w)(t-t')}dt'.
\end{eqnarray} \normalsize
\begin{figure}[t]
\includegraphics[height=5cm,width=8cm]{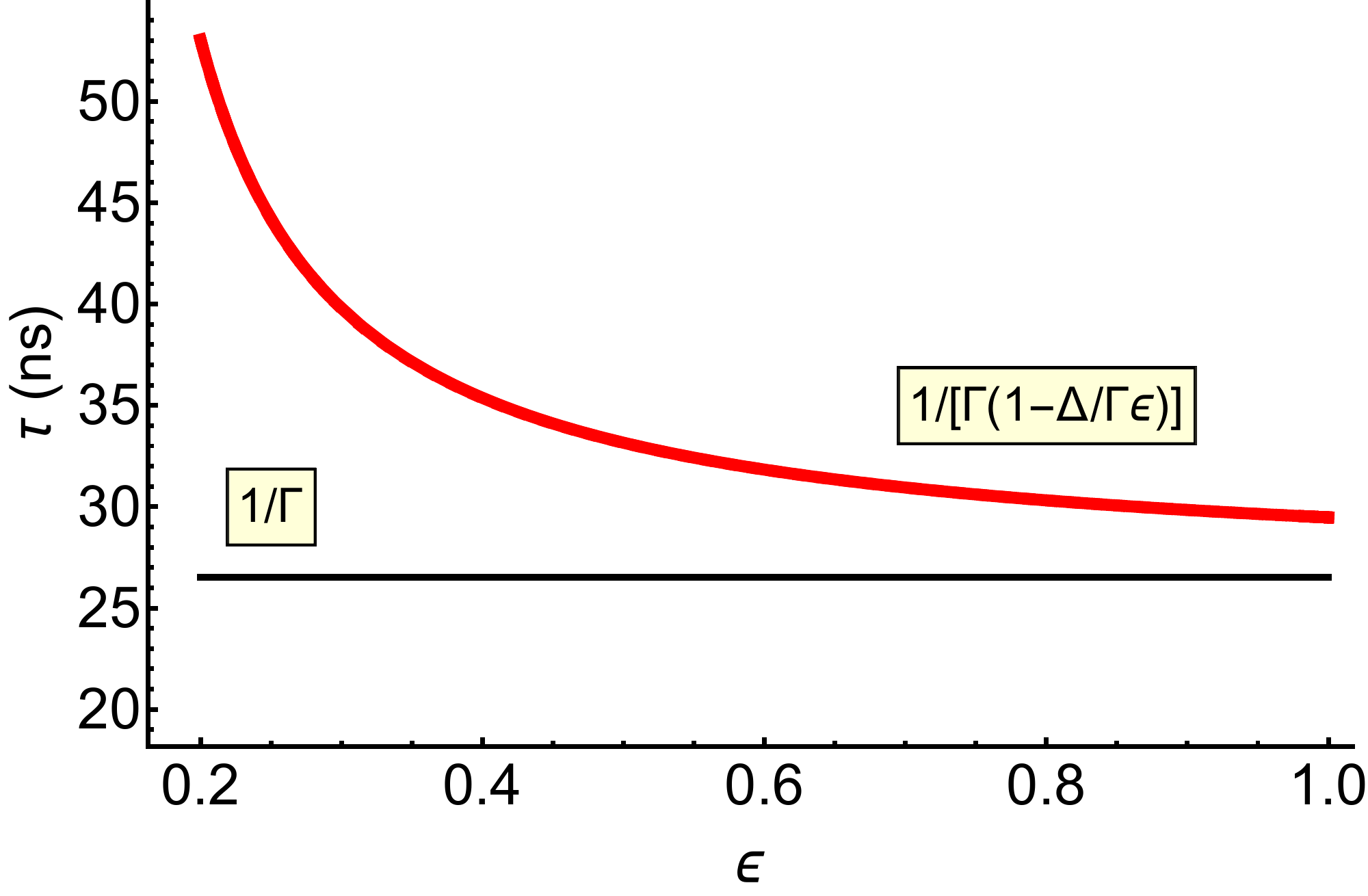}
\caption{\label{fig2}(Color online) The (thick) red line plot shows
increased mean lifetime $\tau$ of atoms as they scatter from an initial state $\vert i \rangle$ to a 
final state $\vert f \rangle$,as a function of $\epsilon$.
Different values of $\epsilon$ denote different final states $\vert f \rangle$.~The (thin) black line shows the 
decay time-scale of the atom to go from $\vert i \rangle$ to the ground state $\vert g \rangle$ in the absence of
post-selection.}
\end{figure}
We assume that the excited state amplitude $\alpha(t)$ varies with a rate $\Gamma \ll w$.~Therefore $\alpha(t)$ changes little and we replace $\alpha(t^{\prime})$ in the 
integrand by $\alpha(t)$.~This is called the Weisskopf-Wigner approximation, 
which can be recognized as a Markov approximation.~Since $\alpha(t)$ varies with a rate $\Gamma \ll w$ the upper bound of the above integral 
can be put to $\infty$, and we have
\small\begin{equation}
\dot{\alpha}(t)= -\dfrac{w^3|\eta|^2}{6\pi\varepsilon_0c^3\hbar}\alpha(t)+\frac{\Delta}{2} \text{cot}(\epsilon) \alpha(t).
\end{equation}\normalsize
By defining $\Gamma=\dfrac{w^3|\eta|^2}{3\pi\varepsilon_0c^3\hbar}$ we have,
 \small\begin{eqnarray}
&&\dot{\alpha}(t)=-\dfrac{\Gamma}{2}\alpha(t)+\frac{\Delta}{2} \text{cot}(\epsilon) \alpha(t).
\end{eqnarray}\normalsize
We use post-selection condition namely $\epsilon \ll 1$, for picking out final states which
are nearly orthogonal to the initial state $\vert i \rangle$, giving
 \small\begin{eqnarray}
&&\dot{\alpha}(t)=-\dfrac{\Gamma}{2}\alpha(t) + \frac{\Delta/\epsilon}{2} \alpha(t), \label{eq:approx} \\
&& \alpha(t) = \alpha(0) e^{(-\Gamma/2)(1-(\Delta/\Gamma \epsilon)) t}.~\label{eq:sol}
\end{eqnarray}\normalsize
Using the fact that at $t = 0, \alpha(0) = 1$, we calculate the probability $|\alpha(t)|^2$ that the atom scatters between the 
desired excited state configurations $\vert i \rangle$
and $\vert f \rangle$. We derive from it the mean time ($\tau$) the atom takes 
for transiting between these two excited state configurations, as a 
function of $\epsilon$ as shown in Fig. \ref{fig2}.~As seen in the figure, for small values of $\epsilon$, that is, 
for nearly orthogonal initial and final
states, the mean-scattering lifetime between these two states increases.~It can even be several times
the mean decay lifetime ($\Gamma$) of the excited state to the ground state.~This increase for scattering between
nearly orthogonal initial and final states is {\color{black} the same as} weak value amplification effect.~Using the values of $\Delta$ and $\Gamma$ pertaining 
to the experiment of  \cite{dayan-prl-111-023604-2013}, we can reproduce remarkably similar mean lifetime values as that obtained by
their measurements. Thus we have rigorously proved that an {\color{black} atomic} excited state interaction with vacuum bath modes 
of the EM field modeled by traditional W-W theory can {\color{black} reproduce all the results of a weak measurements interpretation, provided,
we can} 
\begin{itemize}
\item Define a weak interaction regime which is the regime of elastic resonance scattering.
\item {\color{black} Identify} suitable choices
of initial and final excited states.
\end{itemize}
Our calculation makes transparent the physical principles behind the increased mean lifetime seen in experiments.~{\color{black} In particular,
for the experiment of \cite{dayan-prl-111-023604-2013}, it entirely does away the need for a pointer based weak measurement explanation.\\
\indent Through our analysis we can predict a priori, the maximum value of $\epsilon$, which is an experimentally controllable
parameter, upto which the amplification effect can be obtained.
From Eqn.(\ref{eq:sol}),it is clear that the ratio $\Delta/\Gamma$, which defines the weakness criteria determines this maximum value.
For $\epsilon=\Delta/\Gamma$, the mean scattering time diverges and the values of $\epsilon$ below this value, namely, for
$0<\epsilon<\Delta/\Gamma$ the equation gives unphysical results. This means that for such values of $\epsilon$, scattering from 
$\vert i\rangle$ to $\vert f \rangle$ does not occur.~For accessing smaller values of $\epsilon$ we now need 
to choose a different ratio of $\Delta/\Gamma$.
To emphasize this point, we show
in Fig. 3, a plot of mean scattering lifetime as a function of  $\epsilon$, for values smaller than what is shown in Fig.2. Such values
are made possible by choosing a smaller value for the ratio $\Delta/\Gamma = 0.01$.~From the plot, it is clearly seen that, not only smaller values of epsilon are now 
physically accessible, but also that the corresponding  amplification
effect is much more pronounced than in Fig.2.~This gives an experimenter a powerful choice of parameter values
for observing the amplification effect based on physical reasoning.}
\section{Discussion}
\begin{figure}[t]
\includegraphics[height=5cm,width=8cm]{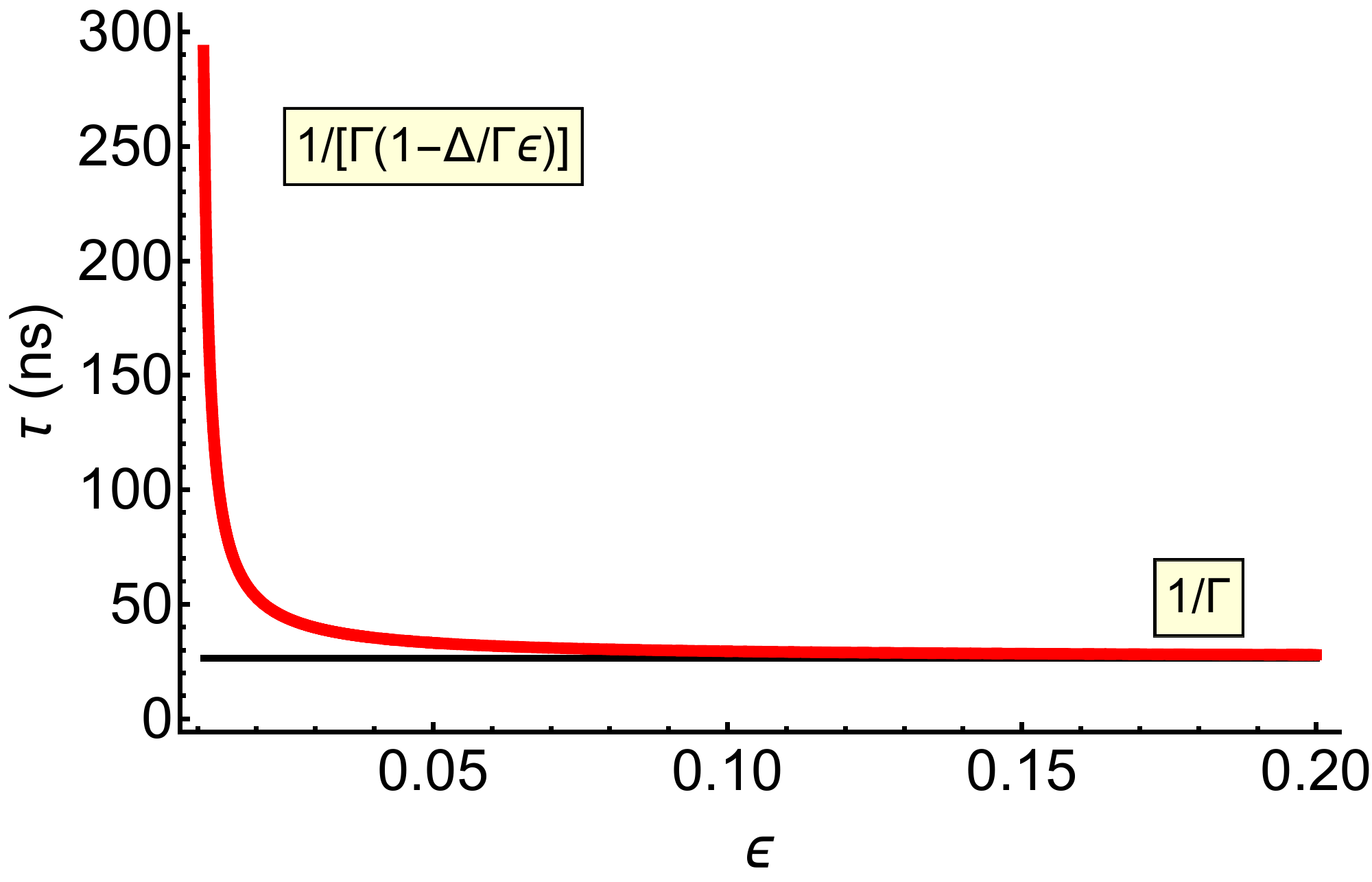}
\caption{\label{fig3}(Color online) The (thick) red line plot shows
increased mean lifetime $\tau$ of atoms as they scatter from an initial state $\vert i \rangle$ to a 
final state $\vert f \rangle$,as a function of $\epsilon$. This plot is for a ratio $\Delta/\Gamma = 0.01$.~The (thin) black line shows the 
decay time-scale of the atom to go from $\vert i \rangle$ to the ground state $\vert g \rangle$ in the absence of
post-selection.~In comparison with Fig.2 which is obtained for a ratio $\Delta/\Gamma = 0.1$, we see that the amplification effect 
is more pronounced for nearly orthogonal final states.}
\end{figure}
{\color{black} For an excited state atom interacting with a bath}, the W-W theory shows that the weak measurement induced by the multitude of vacuum bath modes on the excited
states of the V-system preserves the weak value amplification effect as traditionally defined by \cite{aharanov-prl-60-1351-1988}.~It is interesting 
to note that the interference criteria which is necessary for seeing this effect \cite{sudarshan-prd-40-2112-1989},
now occurs in the interference  arising from the $\vert i \rangle$ and $\vert f \rangle$ interacting with
the very same bath mode at any instant, but averaged over all the modes of the bath.\\
\indent As a corollary of our analysis, we have arrived at a very general result that configurational 
changes within the excited state manifold happen in timescales which are very 
different from the decay time of the excited state to the ground state.~To study such scattering events
between nearly orthogonal excited state configurations, we have to be in the regime of weak, elastic and resonance interaction.~Under such an interaction, we have shown that the scattering time-scale increases as the states
become more and more orthogonal to each other and diverges for states which are exactly orthogonal.~Such a trend of increased mean lifetime is not surprising, 
considering the fact that the fastest timescale in a system-bath
interaction is the decoherence time-scale $1/\Gamma$  \citep{zurek2007decoherence} and any other time-scale in the problem is slower
compared to this decay time.~Within this period $1/\Gamma$, the atom
goes from an initially prepared pure state to a statistical mixture.~Therefore in a traditional non-post-selected experiment there is 
little chance to observe other kind of state changes which does not involve decay to the ground state as the excited state decay occurs more frequently.~However, performing a weak measurement with post-selection makes it possible to isolate events which are rare in occurrence.\\
\indent The W-W theory we have employed is applicable to a wide class of experiments where an excited state 
configuration is interacting with vacuum modes of the EM field.~Thus the theory is applicable for collective excited
states in an atom-atom interaction process.~As a concrete example, we take the case of Forster resonances between excited states of Rydberg atoms
induced by dipole-dipole interaction between the atoms  \citep{nat-phys-10-914,prl-114-113002}.~In a basis
consisting of collective excited states of both atoms and using post-selected calculations as we have done, we can 
evaluate scattering time-scales between specific excited state changes.~This will not only give scattering times between nearly degenerate 
excited states, but more importantly, it will bring out time scales of those
rare scattering events between states which 
are nearly orthogonal.~This is indeed very important because this will
enable evaluation of dipole-dipole scattering coefficient $C_6$ to a greater degree of sensitivity than what is 
currently available. 
\section{Conclusions.}
In resonant elastic scattering, a weak interaction regime between excited states and vacuum bath modes of the EM field
can be established, when the mean separation between the excited states is smaller than the mean decay width 
of the excited states.~In this paper, such an interaction is described for the excited state pair of a V-type
atomic system, using Weiskopff-Wigner (W-W) theory of spontaneous emission.~We show that in this regime, 
an increase in the time-scale of scattering occurs
for events that takes the atom between nearly orthogonal excited state superpositions.~This increased time-scale
is the same as that occurring during a weak value amplification effect.~We show that this effect singles out rare scattering events 
between excited states in a system dominated by frequent decay of the excited state to the ground state.~Thus
for the first time we show that a system-bath interaction can be modeled as a weak interaction 
that can give rise to a corresponding weak value amplification effect.~This opens up a novel and powerful method, to estimate
mean scattering time scales for events which keep the atom within the
excited state manifold.~By applying our theory to rare and long-lived scattering events in the common excited manifold of a 
multi-atom system like that of a Rydberg ensemble we can obtain very sensitive measurements of atom-atom scattering coefficients. 
\bibliography{ref}
\end{document}